\documentclass{aa}

\usepackage{graphicx}
\usepackage{txfonts}
\usepackage{lipsum}
\usepackage{subcaption}         % necessary for continued figures, example in section 3
\usepackage{lscape}             % to rotate a single page table, example in appendix.
\usepackage{placeins}           % useful with \FloatBarrier, to keep
\makeatletter
\let\linenumbers\relax
\makeatother

\begin{document}

   \title{Tomographic Mapping of Primordial Minihalos through Rotational Emission Lines of Molecular Hydrogen}
   \titlerunning{Tomography of Primordial Minihalos in H$_2$ Lines}

   \author{Yu. Kulinich\corrauth{yuriy.kulinich@lnu.edu.ua}
          \and
          B. Novosyadlyj
          }

   \institute{Astronomical Observatory of Ivan Franko National University of Lviv, Kyryla \& Methodia str., 8, Lviv, 79005, Ukraine}

\abstract
  % Context
  {The formation of the first gravitationally bound structures during the Dark Ages and Cosmic Dawn was accompanied by molecular cooling, driven primarily by the rotational transitions of molecular hydrogen ($\mathrm{H}_2$). The cumulative emission from unresolved primordial minihalos is expected to produce a diffuse molecular line-intensity background, potentially providing a new observational probe of early structure formation and radiative feedback.}
  % Aims
  {We investigate the rotational line emission of primordial minihalos by calculating their individual luminosities over a broad range of halo masses and formation redshifts. We further explore the resulting three-dimensional intensity distribution and global spectrum of $\mathrm{H}_2$ rotational emission over the redshift range $6 \le z \le 35$, assessing the prospects for mapping the first gravitationally bound structures through molecular lines.}
  % Methods
  {We used the \texttt{CosmoChem} chemical-kinetic and thermal-balance code to consistently describe the non-equilibrium chemical evolution and rotational excitation of $\mathrm{H}_2$ during the gravitational collapse and virialization of primordial density perturbations. The resulting single-halo luminosities were combined with cosmological simulations performed using \texttt{21cmFAST} to construct three-dimensional minihalo number-density and molecular line-intensity fields.}
  % Results
  {We find that $\mathrm{H}_2$ dominates the molecular cooling emission of primordial minihalos, with the lowest rotational transitions providing the largest contributions. In particular, the ortho-$\mathrm{H}_2$ $J=3\rightarrow1$ line is the dominant transition near the peak of the emission near an observed frequency of $\nu_{\mathrm{obs}}\sim 1.75\ \mathrm{THz}$. Although its spectral shape resembles that of the Cosmic Infrared Background, its amplitude is approximately eight orders of magnitude lower within the adopted homogeneous-halo approximation.}
  % Conclusions
  {Our results establish a framework for predicting the global and spatially varying molecular line emission from primordial minihalos. The homogeneous-halo approximation may substantially underestimate the signal if realistic internal density inhomogeneities enhance $\mathrm{H}_2$ formation, particularly through three-body reactions in sufficiently dense regions. An increase in the line intensity by even two orders of magnitude would bring the signal closer to the sensitivity range of next-generation far-infrared and submillimetre space missions such as FOSSIL. Moreover, the simultaneous observation of several ortho- and para-$\mathrm{H}_2$ rotational transitions could enable multi-line spectral tomography, providing a means of identifying common structures at the same cosmological redshift and potentially reconstructing the spatial distribution of the first gravitationally bound objects.}


   \maketitle
%
%-------------------------------------------------------------------\section{Introduction}
%\lipsum[1]

The formation of the first gravitationally bound structures marked a
fundamental transition in the evolution of the Universe, from the
relatively homogeneous post-recombination epoch to the emergence of
the first stars and galaxies. In the standard $\Lambda$CDM cosmological
model, the first sites of primordial star formation are expected to be
dark matter minihalos with masses of approximately
$10^5-10^7\,M_\odot$, forming at redshifts $z\sim10-30$.
In these objects, the virial temperature of the gas is generally too
low for efficient atomic hydrogen cooling. The contraction of baryonic
matter and the subsequent gravitational collapse are therefore
regulated primarily by molecular cooling, with molecular hydrogen
($\mathrm{H}_2$) acting as the dominant coolant
\citep{BrommCoppiLarson2002,OmukaiNishi1998,
BrommLarson2004,Yoshida2006,Bromm2013}.
The formation and evolution of primordial minihalos are thus closely
connected to the chemical and thermal history of the gas, which
determines its ability to cool, collapse, and eventually form the first
Population III stars.

The chemical and thermal evolution of primordial gas is governed
by a network of reactions involving hydrogen, helium, and their
molecular species. In addition to molecular hydrogen, hydrogen
deuteride (HD) and helium hydride ($\mathrm{HeH}^+$) can contribute
to the thermal evolution under appropriate physical conditions.
The efficiency of molecular cooling and the resulting line
emission depend on the abundances of individual species,
the gas density and temperature, and the populations of
their rotational and vibrational energy levels. These
quantities evolve simultaneously during gravitational
collapse, while the molecular level populations are
determined by radiative and collisional excitation and
de-excitation processes.
In particular,
the formation of $\mathrm{H}_2$ through gas-phase chemical pathways
and its subsequent rotational-line emission play a central role in
regulating the thermal evolution of minihalos and establishing the
conditions for the formation of the first stars
\citep{GalliPalla1998,GalliPalla2002,Lepp2002,
Bovino2014,Nishijima2024}.

The molecular emission of primordial gas may also provide a valuable
observational probe of the early Universe. Molecular transitions have
well-defined rest-frame frequencies, and cosmological redshifting
shifts their emission into the infrared and submillimetre frequency
ranges accessible to space-based observations. This property provides
the basis for line-intensity mapping (LIM), which uses the cumulative
emission of unresolved sources to trace their spatial distribution
without requiring individual objects to be detected. Unlike
conventional galaxy surveys, intensity mapping can therefore probe
populations of faint and distant sources, including primordial
minihalos that are expected to host the earliest stages of star
formation.

The feasibility of using molecular hydrogen emission to map the
pre-reionization Universe was first explored in detail by
\citet{Gong2013}. They calculated the mean intensity and clustering
power spectrum of several rotational and rovibrational $\mathrm{H}_2$
emission lines from high-redshift structures, investigating their
dependence on the properties of the host halos. Their results indicated
that the $0-0\,S(3)$ transition, corresponding to the
$J=5\rightarrow3$ rotational transition and a rest-frame wavelength
of $9.66\,\mu\mathrm{m}$, produces the strongest signal among the
considered lines over the redshift range $10<z<30$.
The potential of molecular line intensity mapping has subsequently
been investigated for hydrogen deuteride (HD), demonstrating that
it may provide an additional tracer of the high-redshift Universe
\citep{Breysse2022}.
Together, these studies demonstrated that primordial
molecular emission could serve as a tracer of the matter
distribution during the Cosmic Dawn and the preceding epochs.
However, reliable predictions of the signal require a
detailed description of the chemical and thermal evolution
of the emitting gas and the excitation of individual
molecular transitions.

The formation, chemical evolution, and radiative properties of the
first molecules in the early Universe have been investigated in a
series of our previous studies
\citep{Novosyadlyj2018,Novosyadlyj2020,Kulinich2020,
Novosyadlyj2022,KulinichNovosyadlyj2025,Kulinich2025}.
These studies addressed the formation and destruction of primordial
molecules in the intergalactic medium and in dark-age halos, as well
as their excitation and radiative signatures. In particular, we
investigated the thermal emission and resonant scattering of CMB
radiation in the rotational lines of $\mathrm{H}_2$ and
$\mathrm{HeH}^+$, and studied the effects of the chemical and
thermal histories of primordial gas on molecular abundances and
emission from forming halos. In our more recent studies, we
developed an extended chemical-kinetic model comprising 166
reactions involving 20 chemical species, and calculated the
rovibrational level populations of $\mathrm{H}_2$, HD, and
$\mathrm{HeH}^+$, together with the resulting spectral distortions
of the cosmic microwave background (CMB). These investigations
provided the basis for studying the molecular signatures of the
early Universe, from the nearly homogeneous intergalactic medium
to primordial halos.

An important development in this direction has been the application
of semi-numerical methods to simulate the spatial distribution of
high-redshift sources and their line emission. In particular, the
LIMFAST framework extends the capabilities of \texttt{21cmFAST}
to model high-redshift galaxy formation and predict intensity maps
in different spectral lines \citep{MasRibas2023,Sun2023}.
Such approaches provide an efficient means of connecting the
distribution of emitting objects to their host dark matter halos
and the underlying large-scale density field.
Applying these methods to primordial minihalos requires
a physically motivated prescription for their molecular
luminosities, based on the non-equilibrium chemical and
thermal evolution of the gas during gravitational collapse
and virialization.

In this work, we extend our previous chemical-kinetic model to
investigate the molecular emission from collapsing primordial
density perturbations and virialized minihalos. Using our
\texttt{CosmoChem} code, which incorporates a network of 166
chemical reactions involving 20 species, we follow the coupled
chemical and thermal evolution of the gas, accounting for
gravitational compression, cosmological expansion, Compton energy
exchange with the CMB, chemical heating and cooling, and molecular
cooling due to $\mathrm{H}_2$, HD, and $\mathrm{HeH}^+$.
Once the perturbation reaches the virial density, the gas
is heated towards the virial temperature according to a
prescribed time-dependent heating rate acting over a
free-fall timescale. We calculate the rotational level
populations of ortho- and para-$\mathrm{H}_2$ up to $J=9$,
and use them to determine the optical depths and line
brightnesses relative to the CMB, which are required to
construct the emission spectrum of an individual halo.
We then calculate the total $\mathrm{H}_2$ line luminosity
of the halo at the stage of maximum emission, together
with the contributions of individual rotational transitions
and the corresponding molecular fraction. These quantities,
including the total luminosity and the fractional
contributions of individual lines, are subsequently used
to construct spatial maps of the molecular emission.
This approach allows us to investigate the dependence of
the molecular emission on halo mass, collapse redshift,
and evolutionary stage, and to provide the physical
quantities required for modelling its spatial distribution.

To model the spatial distribution of the emission, we combine
the calculated single-halo luminosities with cosmological density
and velocity fields generated by the semi-numerical code
\texttt{21cmFAST} \citep{Mesinger2011}. The minihalo abundance in
each computational cell is estimated using the conditional halo
mass function, with the local lower mass threshold determined by
Lyman--Werner feedback and the upper threshold set by efficient
atomic hydrogen cooling. This approach accounts for the spatial
variation of the minihalo population in response to large-scale
density fluctuations and radiative feedback. We use these
distributions to construct three-dimensional intensity maps and
global spectra of the lowest rotational transitions of
para- and ortho-$\mathrm{H}_2$ over $6\leq z\leq35$.
All cosmological simulations were performed assuming a spatially
flat $\Lambda$CDM cosmology with the parameters
$\Omega_{\mathrm{m}}=0.3096$, $\Omega_{\mathrm{b}}=0.0490$,
$h=0.6766$, $\sigma_8=0.8102$, and $n_{\mathrm{s}}=0.9665$.
The simulations were carried out using \texttt{21cmFAST} in a
$(200\ \mathrm{Mpc})^3$ comoving volume, with a high-resolution
density field of $500^3$ cells and an ionization and thermal
grid of $200^3$ cells, corresponding to a spatial resolution
of $1\ \mathrm{Mpc}$ for the latter.

The resulting intensity maps may also provide a basis for future
multi-line spectral tomography of primordial minihalos. Since
different rotational transitions are observed at different
frequencies for the same cosmological redshift, according to
$\nu_{\mathrm{obs}}=\nu_0/(1+z)$,
observations in several molecular lines could probe the same
underlying population of sources at different frequencies.
Cross-correlating such maps may help identify common spatial
structures, distinguish the primordial molecular signal from
foreground emission, and ultimately reconstruct the three-dimensional
distribution of the first gravitationally bound structures.

The paper is organized as follows. In Sect.~\ref{sec:floats},
we describe the chemical and thermal evolution of primordial
gas, the rotational excitation of molecular hydrogen, and
the calculation of individual halo luminosities. In
Sect.~\ref{sec:h2_mapping}, we present the methodology for
constructing three-dimensional $\mathrm{H}_2$ intensity maps
and global line-intensity spectra from the halo luminosities
and cosmological simulations. Finally, in
Sect.~\ref{sec:conclusions}, we discuss the main results,
compare our predictions with previous studies, and consider
the prospects for detecting primordial molecular emission
and applying multi-line spectral tomography to the study
of the first gravitationally bound structures.

\section{The Chemistry and Rotational Excitation of Molecular Hydrogen in Early-Universe Halos} \label{sec:floats}

\subsection{Primordial chemistry in the cosmological background
and homogeneous density perturbations}
\label{sec:primordial_chemistry}

We describe the primordial chemical evolution using the
CosmoChem reaction network, which includes 20 atomic, ionic,
and molecular species and 166 chemical reactions involving
hydrogen, helium, and deuterium. The network accounts for
interactions with CMB photons and free electrons, as well as
two-body reactions between the chemical species. Its composition
and reaction rates are described in detail in
\citet{KulinichNovosyadlyj2025}. The evolution of the fractional
abundance of each species, $x_i=n_i/n_{\rm H}$, is governed by
the system of kinetic equations
\citep{Puy1993,Galli1998,Vonlanthen2009,Novosyadlyj2018}:
\begin{equation}
\begin{aligned}
\frac{dx_i}{dt}
={}&
n_{\rm H}\sum_{m,n} k_{mn}^{(i)}x_mx_n
+n_{\rm H}^2\sum_{m,n,p} k_{mnp}^{(i)}x_mx_nx_p
\\
&+\sum_m k_{m\gamma}^{(i)}x_m
-n_{\rm H}x_i\sum_j k_{ij}x_j
\\
&-n_{\rm H}^2x_i\sum_{j,k}k_{ijk}x_jx_k
-k_{i\gamma}x_i ,
\end{aligned}
\label{eq:chem_network}
\end{equation}
where $k_{mn}^{(i)}$ and $k_{ij}$ are the rate coefficients
for two-body reactions, $k_{mnp}^{(i)}$ and $k_{ijk}$ are
the corresponding coefficients for three-body reactions, while
$k_{m\gamma}^{(i)}$ and $k_{i\gamma}$ describe reactions
involving photons.

We first calculate the chemical evolution in the homogeneous
cosmological background, adopting the standard cosmological
model and the corresponding thermal history of the
post-recombination Universe. The resulting abundances provide
the initial chemical conditions for the subsequent evolution
of primordial density perturbations.

To describe the formation of minihalos, we consider initially
homogeneous spherical density perturbations, within which the
gas density, temperature, and chemical composition are assumed
to be spatially uniform at each evolutionary stage. This
approximation corresponds to the homogeneous top-hat model
of gravitational collapse, which has been employed in our
previous studies of halo formation and primordial chemistry
\citep{Novosyadlyj2018,Novosyadlyj2020}.
The perturbations evolve under their own gravitational
instability, while their mean density increases relative to
the cosmological background. Accordingly, the hydrogen
number density inside a perturbation is given by
\begin{equation}
n_{\rm H}(t)=\bar n_{\rm H}(t)
\left[1+\delta_{\rm b}(t)\right],
\label{eq:nH_pert}
\end{equation}
where $\bar n_{\rm H}(t)$ is the background hydrogen number
density and $\delta_{\rm b}(t)$ is the spatially uniform
baryonic density contrast within the perturbation.

The same chemical reaction network is solved for both the
homogeneous background and the collapsing perturbations.
However, in the latter case, the reaction rates are evaluated
using the time-dependent local gas density and temperature.
The increasing density enhances two-body reactions and
therefore modifies the balance between molecular formation
and destruction compared with the homogeneous cosmological
background.

The formation of molecular hydrogen proceeds predominantly
through two ion-mediated pathways. The first involves the
negative hydrogen ion:
\begin{equation}
\mathrm{H+e^-\rightarrow H^-+\gamma},
\qquad
\mathrm{H^-+H\rightarrow H_2+e^-},
\label{eq:Hminus_channel}
\end{equation}
whereas the second proceeds through the molecular hydrogen ion:
\begin{equation}
\mathrm{H+H^+\rightarrow H_2^++\gamma},
\qquad
\mathrm{H_2^++H\rightarrow H_2+H^+}.
\label{eq:H2plus_channel}
\end{equation}
The relative contribution of these pathways changes with
redshift, with the H$_2^+$ channel being more important at
earlier epochs and the H$^-$ channel dominating at lower
redshifts \citep{KulinichNovosyadlyj2025}.

We follow the coupled chemical and thermal evolution of
homogeneous perturbations from their linear growth stage
until they reach the virial density and form minihalos.
The resulting molecular abundances and gas temperatures
provide the input for calculating the populations of
rotational levels and the luminosities of individual
H$_2$ emission lines. The adopted approach extends our
previous studies of the formation, chemical composition,
and molecular radiation of primordial halos
\citep{Novosyadlyj2018,Novosyadlyj2020,Novosyadlyj2022},
with particular emphasis here on the contribution of
individual H$_2$ rotational transitions.

\subsection{Rotational level populations of molecular hydrogen}
\label{sec:floats}
The populations of the rovibrational levels of primordial molecules
are governed by radiative transitions, collisions with the constituents
of the baryonic gas, and chemical formation and destruction processes.
Their time evolution is described by a system of kinetic equations
that accounts for the corresponding radiative and collisional
transition rates \citep{KulinichNovosyadlyj2025}:
\begin{equation}
\frac{dX_j}{dt}
=
\frac{k_{\rm f}}{n}
(\alpha_j-X_j)
+
\frac{1}{n}
\left(\frac{dn_j}{dt}\right)_{\rm trans},
\label{dX_jdt}
\end{equation}
where $X_j=n_j/n$ is the fractional population of molecular level
$j$, $n_j$ is the number density of molecules in that level, and
$n=\sum_j n_j$ is the total molecular number density. Here,
$k_{\rm f}$ is the total molecular formation rate per unit volume,
and $\alpha_j$ specifies the fraction of newly formed molecules
initially produced in level $j$, with $\sum_j\alpha_j=1$.
The first term describes the effect of chemical formation and
destruction on the level populations, under the assumption that
the destruction rate is independent of the internal molecular
state. The second term accounts for transitions between molecular
levels induced by radiative and collisional processes:
\begin{equation}
\left(\frac{dn_j}{dt}\right)_{\rm trans}
=
\sum_{i\ne j}
\left(
n_i R_{ij}-n_j R_{ji}
\right),
\end{equation}
where $R_{ij}$ is the total transition rate per molecule from
level $i$ to level $j$. It comprises radiative and collisional
contributions,
\begin{equation}
R_{ij}=R_{ij}^{\rm \gamma}+R_{ij}^{\rm col}.
\end{equation}
The collisional contribution is given by
\begin{equation}
R_{ij}^{\rm col}
=
\sum_c n_c C_{ij}^{(c)}(T_{\rm m}),
\end{equation}
where $n_c$ is the number density of collision partner $c$, and
$C_{ij}^{(c)}(T_{\rm m})$ is the corresponding temperature-dependent
rate coefficient. The radiative rates are determined by the
Einstein coefficients and the spectral energy density of the
ambient radiation field. This kinetic description permits
departures from thermal equilibrium, since the level populations
depend on the competition between radiative and collisional
transitions and the distribution of newly formed molecules among
the internal states.

Molecular hydrogen is a homonuclear diatomic molecule with no
permanent electric dipole moment. Its rotational radiative
transitions therefore proceed through electric quadrupole emission
and obey the selection rule $\Delta J=\pm2$. The rotational levels
belong to two distinct nuclear-spin species: ortho-H$_2$, with odd
rotational quantum numbers $J$, and para-H$_2$, with even $J$.
Their nuclear-spin statistical weights are 3 and 1, respectively.
Since radiative transitions preserve nuclear-spin symmetry, they
do not convert ortho-H$_2$ into para-H$_2$ or vice versa.

In this work, we consider rotational transitions within the ground
vibrational state ($v=0$), focusing in particular on the
$J=2\rightarrow0$ transition of para-H$_2$ and the
$J=3\rightarrow1$ transition of ortho-H$_2$, as well as the
higher-order transitions $J\rightarrow J-2$. The spontaneous
emission coefficients and transition frequencies for the lowest
rotational levels are taken from \citet{Roueff:2019}. We also
account for the interaction of H$_2$ with the cosmic microwave
background (CMB). For a transition from an upper level $u$ to a
lower level $l$, the radiative de-excitation and excitation rates,
including spontaneous emission, stimulated emission, and
absorption, are
\begin{equation}
R_{ul}^{\gamma}
=
A_{ul}+B_{ul}U(\nu_{ul}),
\qquad
R_{lu}^{\gamma}
=
B_{lu}U(\nu_{ul}),
\end{equation}
where $U(\nu_{ul})$ is the CMB spectral energy density evaluated
at the transition frequency $\nu_{ul}$.

The rotational populations are also affected by collisions with
neutral hydrogen and helium atoms and protons. The collisional
de-excitation rate coefficients for H$_2$--H and H$_2$--He are
adopted from \citet{Lique:2015} and \citet{Nolte_2012},
respectively. The corresponding excitation coefficients are
obtained from the principle of detailed balance:
\begin{equation}
C_{lu}(T_{\rm m})=
\frac{g_u}{g_l}
\exp\left[
-\frac{E_u-E_l}{k_{\rm B}T_{\rm m}}
\right]
C_{ul}(T_{\rm m}),
\end{equation}
where $g_u$ and $g_l$ are the statistical weights of the upper
and lower rotational levels, respectively, and $T_{\rm m}$ is
the kinetic temperature of the gas.

Although protons are less abundant than neutral hydrogen and
helium, collisions with protons are particularly important because
they can induce transitions between ortho- and para-H$_2$.
We account for these processes using the rate coefficients from
\citet{Gerlich:1990}. Thus, the rotational excitation and
de-excitation of H$_2$ are determined by the combined action of
the CMB radiation field and collisions with the main components
of the primordial gas. The resulting level populations are used
to calculate the luminosities of individual H$_2$ rotational
lines.

\subsection{Evolution of the Gas Temperature}
\label{sec:gas_temperature}

\begin{figure}
   \centering
   \includegraphics[width=0.5\textwidth]{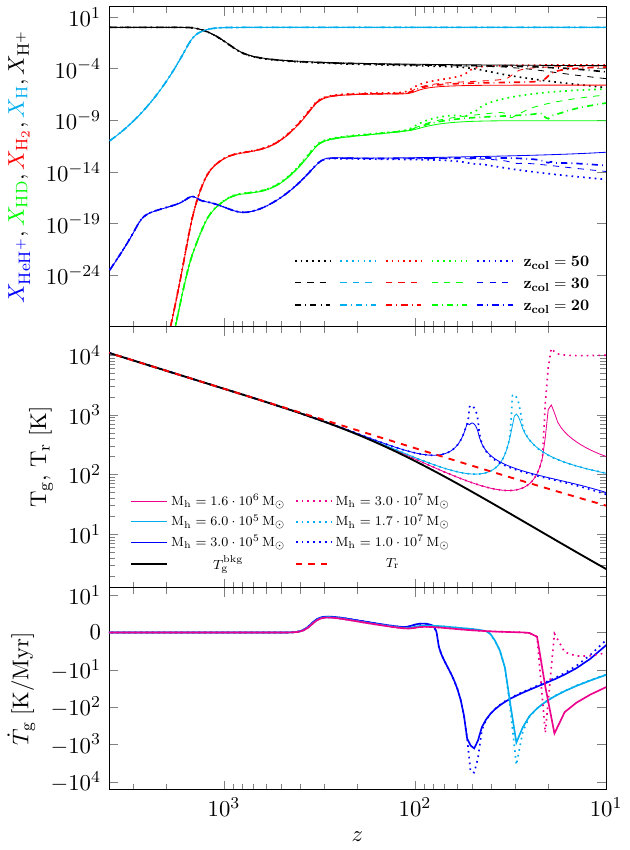}
   \caption{Kinetics of primordial molecule formation (top panel), evolution of the gas temperature (middle panel), and contribution of $\mathrm{H}_2$ rotational transitions to the total gas cooling rate (bottom panel) for three perturbations collapsing into halos at redshifts \(z_{\rm col}=20,\ 30,\) and \(50\).} \label{fig:halo_chem_temp}
\end{figure}

The thermal evolution of the primordial gas is governed by
adiabatic expansion and compression, radiative heating and
cooling, chemical reactions, and energy exchange with the
cosmic microwave background (CMB). The corresponding
temperature evolution equation can be written in the general
form
\begin{equation}
\begin{aligned}
\frac{dT_{\rm g}}{dt} ={}&
\frac{2T_{\rm g}}{3n}
\frac{dn}{dt}
+\frac{2(\Gamma-\Lambda)}
{3k_{\rm B}n_{\rm tot}}\\
&+\frac{T_{\rm g}}{\mu}\frac{d\mu}{dt}
+\frac{8\sigma_{\rm T}a_{\rm r}T_\gamma^4}
{3m_{\rm e}c}
\frac{x_{\rm e}}{1+x_{\rm e}+f_{\rm He}}
(T_\gamma-T_{\rm g}),
\end{aligned}
\label{eq:gas_temperature_evolution}
\end{equation}
where $T_{\rm g}$ is the gas kinetic temperature, $n$ is the
number density of hydrogen nuclei, $n_{\rm tot}$ is the total
number density of gas particles, and $\mu$ is the mean molecular
weight in units of the proton mass. The quantities $x_{\rm e}$
and $f_{\rm He}$ denote the free-electron abundance and the
helium abundance relative to hydrogen, respectively.
Furthermore, $\sigma_{\rm T}$, $a_{\rm r}$, and $m_{\rm e}$
are the Thomson scattering cross-section, radiation constant,
and electron mass.

The first term describes adiabatic heating or cooling resulting
from changes in the gas density. Before virialization, the
hydrogen number density evolves due to both cosmological
expansion and the growth of the density perturbation:
\begin{equation}
n(z)=\bar n_0(1+z)^3[1+\delta(z)],
\end{equation}
where $\bar n_0$ is the present-day mean hydrogen number density
and $\delta(z)$ is the density contrast. Accordingly, the
adiabatic term can be expressed as
\begin{equation}
\frac{2T_{\rm g}}{3n}\frac{dn}{dt}
=-2H_{\rm eff}(z)T_{\rm g},
\end{equation}
where the effective expansion rate is defined as
\begin{equation}
H_{\rm eff}(z)=H(z)-
\frac{1}{3}\frac{\dot{\delta}(z)}{1+\delta(z)}.
\label{Heff}
\end{equation}
Here, $H(z)$ is the cosmological expansion rate and the dot
denotes differentiation with respect to cosmic time. The
effective expansion rate thus accounts for both the Hubble
expansion and the gravitational contraction of the perturbation.

When the perturbation reaches the virial density, we assume
that it becomes gravitationally bound and undergoes further
heating towards the virial temperature of the halo. During
this stage, we prescribe a smoothly varying heating rate
over one free-fall time, with its mean value given by
\begin{equation}
\left\langle\frac{dT_{\rm g}}{dt}\right\rangle
=
\frac{T_{\rm vir}(M_{\rm h})-T_{\rm g}(t_{\rm vir})}
{t_{\rm ff}},
\label{eq:mean_heating_rate}
\end{equation}
where $T_{\rm g}(t_{\rm vir})$ is the gas temperature at
virialization. This prescription sets the mean rate of
temperature increase required to bring the gas close to
the virial temperature within one free-fall time. The actual
temperature evolution is determined by the combined effects
of the prescribed heating and radiative and chemical cooling
processes. Consequently, the gas temperature at the end of
this interval may differ slightly from the virial temperature.

The free-fall time, evaluated at the virial density, is
\begin{equation}
t_{\rm ff} =
\sqrt{\frac{3\pi}{32G\rho_{\rm vir}}},
\label{eq:free_fall_time}
\end{equation}
where $\rho_{\rm vir}$ is the mean total mass density of the
virialized halo. The virial temperature is estimated as
\begin{equation}
T_{\rm vir}(M_{\rm h}) =
2000\,{\rm K}
\left(\frac{M_{\rm h}}{10^6\,M_\odot}\right)^{2/3}
\left(\frac{1+z_{\rm col}}{20}\right),
\label{eq:virial_temperature}
\end{equation}
where $M_{\rm h}$ is the halo mass and $z_{\rm col}$ is
the collapse redshift. During the subsequent evolution,
the gas temperature is governed by the thermal evolution
equation, with $T_{\rm vir}$ serving as the characteristic
temperature scale of the halo.

The second term in Eq.~(\ref{eq:gas_temperature_evolution})
describes the net thermal energy exchange due to heating and
cooling processes, where $\Gamma$ and $\Lambda$ are the total
volumetric heating and cooling rates, respectively. The third
term accounts for changes in the mean molecular weight caused
by the evolution of the chemical composition. The last term
describes Compton energy exchange between the gas and the CMB
through scattering of CMB photons by free electrons, with
$T_\gamma=T_{\rm CMB}(z)$.

The net heating and cooling rate includes contributions from
molecular excitation and de-excitation, chemical reactions,
and other relevant physical processes. The molecular
contribution arises from collisional transitions between
rotational and vibrational energy levels of primordial
molecules and can be written as
\begin{equation}
(\Gamma-\Lambda)_{\rm mol}
=
\sum_c\sum_j\sum_{j'}
n_j n_c k^{(c)}_{jj'}
(E_j-E_{j'}),
\label{eq:molecular_heating_cooling}
\end{equation}
where $n_j$ is the number density of molecules in the initial
quantum state $j$, $n_c$ is the number density of the
collision partner $c$, and $k^{(c)}_{jj'}$ is the corresponding
temperature-dependent collisional transition rate coefficient.
The quantities $E_j$ and $E_{j'}$ are the energies of the
initial and final molecular states, respectively. The
summation is performed over all relevant collision partners
and molecular transitions. With this sign convention,
collisional de-excitation ($E_j>E_{j'}$) heats the gas,
whereas collisional excitation ($E_j<E_{j'}$) cools it.

The chemical contribution accounts for the energy released
or absorbed during the formation and destruction of primordial
species. It can be expressed as
\begin{equation}
(\Gamma-\Lambda)_{\rm chem}
=
\sum_i R_i(T_{\rm g})E_i,
\label{eq:chemical_heating_cooling}
\end{equation}
where $R_i$ is the volumetric rate of the $i$-th chemical
reaction and $E_i$ is the net thermal energy deposited into
or extracted from the gas per reaction. Positive and negative
values of $E_i$ correspond to chemical heating and cooling,
respectively. For a two-body reaction involving reactants
$a_i$ and $b_i$, the reaction rate is
\begin{equation}
R_i=k_i(T_{\rm g})n_{a_i}n_{b_i},
\end{equation}
with the appropriate statistical factor for identical
reactants included according to the adopted definition of
the rate coefficient.

The total net heating and cooling rate is therefore
decomposed as
\begin{equation}
\Gamma-\Lambda =
(\Gamma-\Lambda)_{\rm mol}
+(\Gamma-\Lambda)_{\rm chem}
+(\Gamma-\Lambda)_{\rm other},
\label{eq:total_heating_cooling}
\end{equation}
where the last term accounts for other heating and cooling
processes included in the thermal model.

In the post-virialization stage, the mean density of the gravitationally bound perturbation is assumed to remain constant, so the density-dependent adiabatic term in Eq.~(\ref{eq:gas_temperature_evolution}) vanishes. The gas temperature then increases at the mean rate given by Eq.~(\ref{eq:mean_heating_rate}) to capture the relaxation toward the virial temperature over a free-fall timescale. Throughout this stage, chemical abundances and molecular level populations continue to evolve, enabling self-consistent calculation of the molecular line emission and cooling rates along the adopted thermal trajectory.

Figure~\ref{fig:halo_chem_temp} shows the formation kinetics of primordial molecules (top panel), the evolution of the gas temperature (middle panel), and the contribution of $\mathrm{H}_2$ rotational transitions to the total gas cooling rate (bottom panel) for three perturbations collapsing into haloes at redshifts $z_{\rm col} = 20$, $30$, and $50$.

Our calculations reproduce the expected asymptotic behavior of the molecular hydrogen fraction $f_{\mathrm{H}_2} \equiv n_{\mathrm{H}_2} / n_{\mathrm{H}}$  in haloes. In dust-free primordial gas without a strong external dissociating radiation field, $f_{\mathrm{H}_2}$ approaches a quasi-steady saturation level during recombination and collapse that scales as a power law in the virial temperature $T_{\mathrm{vir}}$:
\begin{equation}
f_{\mathrm{H}_2,\mathrm{sat}} \approx (3 - 5) \times 10^{-4} \left( \frac{T_{\mathrm{vir}}}{1000\text{ K}} \right)^{\alpha},
\label{eq:fh2_sat}
\end{equation}
with $\alpha \approx 1.52 - 1.6$ across typical minihalo temperatures ($100\text{ K} \lesssim T_{\mathrm{vir}} \lesssim 5000\text{ K}$) \citep[e.g.,][]{Nebrin2023}. This power-law scaling reflects the thermal sensitivity of $\mathrm{H}_2$ formation via the $\mathrm{H}^-$ channel ($\mathrm{e}^- + \mathrm{H} \rightarrow \mathrm{H}^- + \gamma$ followed by $\mathrm{H}^- + \mathrm{H} \rightarrow \mathrm{H}_2 + \mathrm{e}^-$) prior to the onset of strong collisional dissociation or atomic cooling \citep[e.g.,][]{Tegmark1997,GalliPalla1998,GloverAbel2008}.
At the moment of collapse and virialization, $z_{coll}$, the fraction of hydrogen molecules in the halo does not instantly reach a saturated level. In more massive haloes high virial temperatures suppress $f_{\mathrm{H}_2}$ at higher collapse redshifts $z_{\mathrm{col}}$ in two stages:
\begin{enumerate}
    \item \textit{Collisional dissociation:} Above $T_{\mathrm{diss}} \approx 3350\text{ K}$, reactions such as $\mathrm{H}_2 + \mathrm{H} \rightarrow 3\mathrm{H}$ and $\mathrm{H}_2 + \mathrm{e}^- \rightarrow 2\mathrm{H} + \mathrm{e}^-$ rapidly destroy $\mathrm{H}_2$ until a residual chemical balance is reached.
    \item \textit{Collisional ionization drop:} Near $T_{\mathrm{ion}} \approx 1.05 \times 10^4\text{ K}$, collisional ionization of atomic hydrogen boosts free electron and proton abundances, accelerating destruction via charge transfer ($\mathrm{H}_2 + \mathrm{H}^+ \rightarrow \mathrm{H}_2^+ + \mathrm{H}$) and lowering the residual fraction by roughly two more orders of magnitude.
\end{enumerate}

Combining the power-law trend with redshift scaling and thermal suppression $S(T_{\mathrm{vir}})$, we approximate the mass-averaged molecular fraction at $z_{col}$ as
\begin{equation}
f_{\mathrm{H}_2}(M_{\mathrm{halo}}, z_{\mathrm{col}}) = f_0 \left( \frac{T_{\mathrm{vir}}}{1000\text{ K}} \right)^{1.52} \left( \frac{1 + z_{\mathrm{col}}}{11.68} \right)^{0.6} S(T_{\mathrm{vir}}),
\label{eq:fh2_suppressed}
\end{equation}
where $f_0 \approx 1.1 \times 10^{-5}$ is the normalization at $T_{\mathrm{vir}} = 1000\text{ K}$ and
\begin{equation}
S(T_{\mathrm{vir}}) = \left[1 - S_{\min}(T_{\mathrm{vir}})\right] \exp\left[ -\left( \frac{T_{\mathrm{vir}}}{T_{\mathrm{diss}}} \right)^{\beta} \right] + S_{\min}(T_{\mathrm{vir}}),
\label{eq:suppression_factor}
\end{equation}
with steepness index $\beta \approx 3.6$. The residual floor is given by
\begin{equation}
S_{\min}(T_{\mathrm{vir}}) = 1.5 \times 10^{-5} \left( \frac{T_{\mathrm{vir}}}{8000\text{ K}} \right)^{-3} \left[ 1 + \left( \frac{T_{\mathrm{vir}}}{T_{\mathrm{ion}}} \right)^{20} \right]^{-1},
\label{eq:s_min_floor}
\end{equation}
capturing both high-temperature chemical equilibrium and the sharp drop above $T_{\mathrm{ion}}$.

\subsection{Halo Line Brightness Contrast against the CMB}
\label{sec:halo_brightness}

The populations of the rovibrational levels of primordial
molecules begin to deviate from radiative equilibrium at
redshifts $z\lesssim 300$, as collisions with other components
of the baryonic gas become increasingly important. At these
epochs, the gas kinetic temperature is slightly lower than
the CMB temperature, and collisional processes modify the
molecular level populations relative to their radiative
equilibrium values. These departures can be characterized
by the excitation temperature of a transition $u\rightarrow l$,
defined as
\begin{equation}
T_{\rm ex}^{ul} =
\frac{E_u-E_l}{k_{\rm B}}
\left[
\ln\left(\frac{g_u n_l}{g_l n_u}\right)
\right]^{-1},
\label{eq:excitation_temperature}
\end{equation}
where $E_u$ and $E_l$ are the energies of the upper and lower
levels, $n_u$ and $n_l$ are their number densities, and
$g_u$ and $g_l$ are the corresponding statistical weights.

The brightness contrast of a molecular line against the CMB
background is determined by the difference between the line
source function and the background radiation intensity.
For a homogeneous medium with a constant excitation temperature,
the differential specific intensity in the rest frame of the
emitting gas is
\begin{equation}
\delta I_{ul} =
\left[
B_{\nu_{ul}}(T_{\rm ex}^{ul})
-
B_{\nu_{ul}}(T_R)
\right]
\left(1-e^{-\tau_{ul}}\right),
\label{eq:line_brightness_contrast}
\end{equation}
where $B_\nu(T)$ is the Planck function,
$T_R=T_{\rm CMB}(z)$ is the radiation temperature at the
redshift of the emitting region, and $\tau_{ul}$ is the
optical depth of the transition. In the optically thin limit,
$\tau_{ul}\ll 1$, Eq.~(\ref{eq:line_brightness_contrast})
reduces to
\begin{equation}
\begin{aligned}
\delta I_{ul}\simeq
\frac{2h\nu_{ul}^{3}}{c^2}
\bigg[
&\frac{1}{\exp\left(h\nu_{ul}/k_{\rm B}T_{\rm ex}^{ul}\right)-1}\\
-&\frac{1}{\exp\left(h\nu_{ul}/k_{\rm B}T_R\right)-1}
\bigg]\tau_{ul}.
\end{aligned}
\label{eq:line_brightness_thin}
\end{equation}
Thus, a transition appears in emission when
$T_{\rm ex}^{ul}>T_R$ and in absorption when
$T_{\rm ex}^{ul}<T_R$.

The optical depth of a molecular transition depends on the
physical conditions of the gas and its dynamical state.
During the cosmological growth of a density perturbation,
before virialization, we adopt the narrow-line approximation,
in which the effective velocity gradient is determined by
the combined effects of cosmic expansion and gravitational
contraction. For a homogeneous spherical perturbation, this
gradient is given by $H_{\rm eff}(z)$ from
Eq.~(\ref{Heff}). The corresponding optical depth is
\begin{equation}
\tau_{ul}(z)=
\frac{\lambda_{ul}^{3}A_{ul}}
{8\pi |H_{\rm eff}(z)|}
n_{\rm mol}(z)
\left[
\frac{g_u}{g_l}X_l(z)-X_u(z)
\right],
\label{eq:optical_depth_cosmological}
\end{equation}
where $\lambda_{ul}=c/\nu_{ul}$, $A_{ul}$ is the Einstein
coefficient for spontaneous emission, $n_{\rm mol}$ is the
total molecular number density, and $X_l$ and $X_u$ are
the fractional populations of the lower and upper levels,
respectively. The absolute value of the velocity gradient
is used to define the positive optical-depth scale for
either expansion or contraction, provided there is no
population inversion.

Once the perturbation reaches the virial density, its
gravitational collapse is assumed to be halted and the gas
becomes gravitationally bound within the halo. In this regime,
the optical depth is determined by the finite spatial extent
of the halo and the thermal Doppler broadening of the molecular
line. Assuming a homogeneous gas distribution and neglecting
systematic internal velocity gradients and non-thermal
broadening, the line-centre optical depth along the central
line of sight is
\begin{equation}
\tau_{ul}^{\rm vir}(0)=
\frac{A_{ul}\lambda_{ul}^{3}}
{8\pi^{3/2}}
\frac{2R_{\rm vir}}{v_{\rm th}}
n_{\rm mol}
\left[
\frac{g_u}{g_l}X_l-X_u
\right],
\label{eq:optical_depth_virial}
\end{equation}
where $R_{\rm vir}$ is the virial radius and
\begin{equation}
v_{\rm th}=
\sqrt{\frac{2k_{\rm B}T_{\rm g}}{m_{\rm mol}}}
\label{eq:thermal_doppler_velocity}
\end{equation}
is the thermal Doppler velocity, with $m_{\rm mol}$ being
the molecular mass.

The two expressions for the optical depth describe distinct
physical regimes. Before virialization, the resonance
interaction is governed by the velocity gradient associated
with the cosmological expansion and the growth of the
perturbation. After virialization, it is determined by the
finite path length through the halo and the thermal line
width. In both regimes, the molecular abundances and level
populations are obtained from the corresponding chemical
and thermal evolution of the gas.

\begin{figure}
   \centering
   \includegraphics[width=0.46\textwidth]{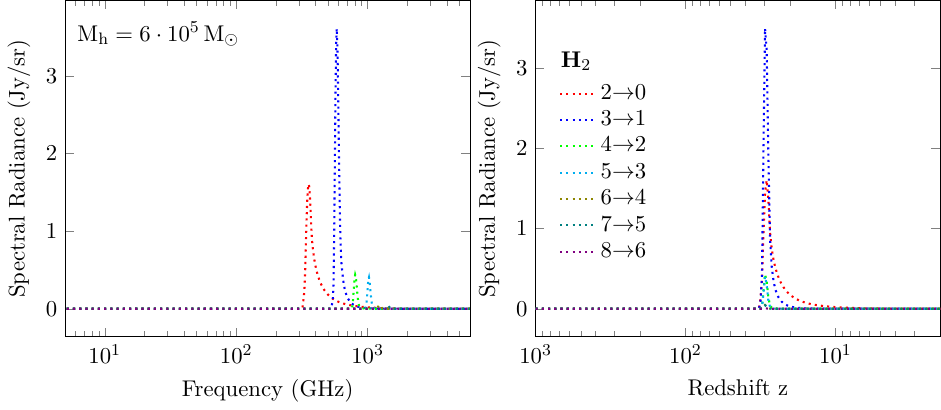}
   \caption{Spectral radiance of individual H\(_2\) rotational transitions in a minihalo of mass \(M_h=6\times10^5\,M_\odot\), collapsing at \(z_{\rm col}=30\), as a function of observed frequency (left panel) and emission redshift (right panel).} \label{fig:halo_br_30m6}
\end{figure}

\subsection{Intrinsic luminosity of H$_2$ rotational transitions}

\begin{figure}
   \centering
   \includegraphics[width=0.5\textwidth]{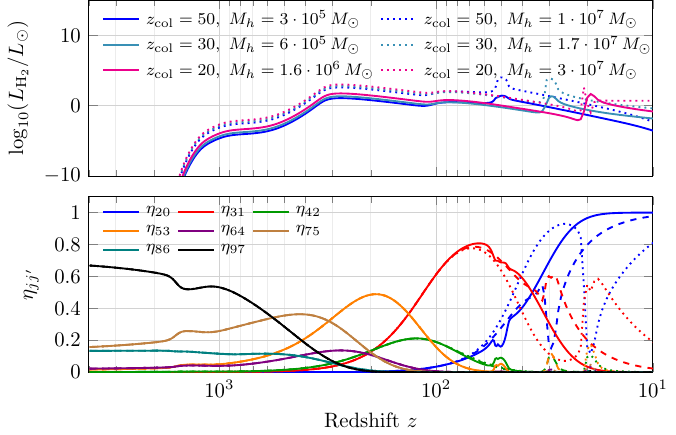}
   \caption{Total H\(_2\) luminosity (top panel) and fractional contributions of individual rotational transitions to the total H\(_2\) line luminosity (bottom panel) as functions of redshift for minihalos of different masses and collapse redshifts.} \label{fig:halo_luminosity}
\end{figure}

The intrinsic luminosity of a molecular hydrogen rotational transition
$j\rightarrow j'$, with $j>j'$, can be estimated from the spontaneous
emission rate of molecules occupying the upper rotational level.
Assuming that the molecular abundance and rotational level populations
are spatially uniform, the emitted line luminosity is given by
\begin{equation}
L_{jj'}^{\rm em}
=
\frac{h\nu_{jj'}A_{jj'}}{m_{\rm H}}\,
X_{\rm H} f_b M_{\rm halo}\,
f_{\rm H_2} X_j .
\label{eq:halo_line_luminosity}
\end{equation}
Here, $h$ is the Planck constant, while $\nu_{jj'}$ and $A_{jj'}$ are
the frequency and Einstein coefficient for spontaneous emission of
the $j\rightarrow j'$ transition, respectively. The hydrogen-atom
mass is denoted by $m_{\rm H}$, and $M_{\rm halo}$ is the total halo
mass. We assume that the gas mass is
$M_{\rm gas}=f_b M_{\rm halo}$, where
$f_b=\Omega_b/\Omega_m$ is the cosmic baryon fraction, and $X_{\rm H}$
is the hydrogen mass fraction of the gas. The molecular abundance is
defined as $f_{\rm H_2}=n_{\rm H_2}/n_{\rm H}$, where $n_{\rm H}$ is
the number density of hydrogen nuclei, and $X_j$ denotes the fraction
of H$_2$ molecules occupying rotational level $j$. Thus, the
abundance and level populations entering
Eq.~(\ref{eq:halo_line_luminosity}) are understood as mass-averaged
quantities over the halo. This expression gives the intrinsic
luminosity produced by spontaneous emission, without accounting for
line absorption, radiative transfer effects, or interactions with
the cosmic microwave background.

To evaluate the rotational level populations \(X_j\) and calculate the total molecular line luminosity, we must account for the substantial increase in the molecular hydrogen abundance during halo collapse and virialization, as illustrated in Fig.~\ref{fig:halo_chem_temp}. Newly formed H\(_2\) molecules are produced predominantly through gas-phase chemical reactions, including the ionic formation channel, with an intrinsic ortho-to-para production ratio of \(3:1\). The rotational populations of these nascent molecules are therefore determined by the state of the reaction products and the conditions prevailing at the time of their formation. In the kinetic rate equation (\ref{dX_jdt}), this effect is explicitly described by the level-dependent formation rate coefficients \(\alpha_j\). As a first approximation, we assume that the rotational populations of newly formed molecules follow a Boltzmann distribution at the local gas temperature \(T_{\rm g}\), with the appropriate nuclear-spin statistical weights. Thus, the formation fractions are given by
\begin{equation}
\alpha_j(T_{\rm g}) =
\frac{g_j \exp\left(-E_j/k_B T_{\rm g}\right)}
{Z(T_{\rm g})},
\label{eq:level_population}
\end{equation}
where \(E_j\) is the rotational excitation energy of level \(j\), \(k_B\) is the Boltzmann constant, and \(g_j\) includes the rotational degeneracy and nuclear-spin statistical weight: \(g_j=2j+1\) for para-H\(_2\) (even \(j\)) and \(g_j=3(2j+1)\) for ortho-H\(_2\) (odd \(j\)), and $Z(T_{\rm g}) = \sum_j g_j \exp\left(-E_j/k_B T_{\rm g}\right)$.

\begin{figure}
   \centering
   \includegraphics[width=0.48\textwidth]{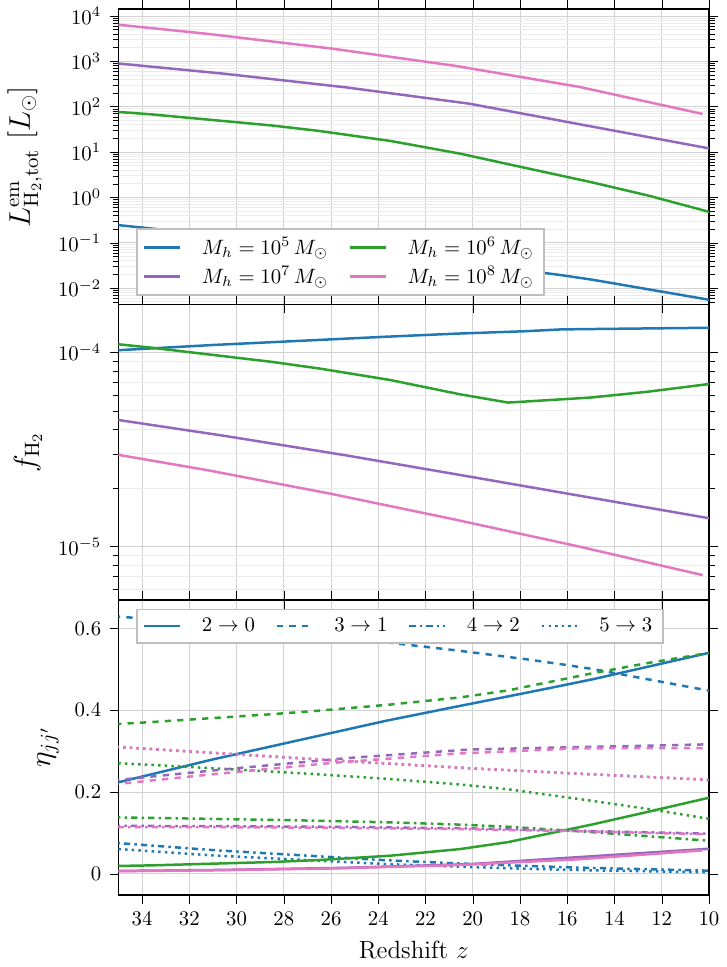}
   \caption{Halo luminosity in molecular hydrogen spectral lines (upper panel), molecular hydrogen fraction (middle panel), and the fraction of the total H$_2$ luminosity contributed by the four lowest rotational transitions (lower panel), for halos of different masses at the epochs corresponding to their peak luminosity, characterized by different redshifts \(z\).}
   \label{fig:hlum}
\end{figure}

The total intrinsic rotational luminosity of molecular hydrogen is
obtained by summing the contributions of all radiative rotational
transitions:
\begin{equation}
L_{\rm H_2, tot}^{\rm em}
=
\sum_{j>j'} L_{jj'}^{\rm em}.
\label{eq:total_H2_luminosity}
\end{equation}
It is therefore convenient to introduce the fractional contribution
of an individual transition to the total H$_2$ rotational luminosity:
\begin{equation}
L_{jj'}^{\rm em}
=
\eta_{jj'}L_{\rm H_2}^{\rm em},
\end{equation}
where
\begin{equation}
\eta_{jj'}=
\frac{h\nu_{jj'}A_{jj'}X_j}
{\displaystyle\sum_{k>k'}h\nu_{kk'}A_{kk'}X_k}.
\label{eq:line_luminosity_fraction}
\end{equation}
Here, $\eta_{jj'}$ characterizes the relative importance of the
$j\rightarrow j'$ transition in the total rotational emission.
By construction, the sum of these fractions over all included
transitions is unity:
\begin{equation}
\sum_{j>j'}\eta_{jj'}=1.
\end{equation}

Figure~\ref{fig:halo_luminosity} shows the evolution of the total H\(_2\) rotational line luminosity and its distribution among individual transitions. The upper panel presents the total rotational line luminosity of homogeneous perturbations that evolve into halos of different masses at different collapse redshifts, while the lower panel shows the fractional contributions, \(\eta_{jj'}\), of individual rotational transitions to the total luminosity. In the upper panel,
solid and dotted curves correspond to the lower- and higher-mass
minihalos, respectively. For collapse redshifts
$z_{\rm col}=50$, 30, and 20, the corresponding halo masses are
$3\times10^5\,M_\odot$ and $10^7\,M_\odot$,
$6\times10^5\,M_\odot$ and $1.7\times10^7\,M_\odot$, and
$1.6\times10^6\,M_\odot$ and $3\times10^7\,M_\odot$,
respectively. In the lower panel, solid, dashed, and dash-dotted
curves represent the lower-mass minihalo models with
$M_{\rm halo}=3\times10^5\,M_\odot$,
$6\times10^5\,M_\odot$, and $1.6\times10^6\,M_\odot$,
corresponding to $z_{\rm col}=50$, 30, and 20, respectively.
As seen in the Figure~\ref{fig:halo_luminosity}, ortho-H\(_2\) dominates during the initial stages of evolution because molecular hydrogen formation produces ortho-H\(_2\) approximately three times more efficiently than para-H\(_2\). During the subsequent expansion and cooling of the Universe, the ortho-H\(_2\) fraction gradually decreases, while the para-H\(_2\) fraction increases \citep{KulinichNovosyadlyj2025}. This is because the lowest rotational level of para-H\(_2\) has a lower energy than the lowest rotational level of ortho-H\(_2\). Conversion between the ortho and para forms can occur through collisions that induce changes in the nuclear-spin state of the molecule. The efficiency of this process is determined by the rates of the corresponding reactions and depends on the physical conditions in the gas. During gravitational collapse, the gas within the perturbation becomes denser and heats up, accompanied by the rapid formation of new H\(_2\) molecules. Since ortho-H\(_2\) is preferentially produced during molecular formation, with an ortho-to-para ratio of approximately 3:1, its fraction increases again. Consequently, the ortho-H\(_2\) line emission from the halo once again becomes dominant during the collapse phase, even if para-H\(_2\) was more abundant in the gas immediately before the onset of collapse.

Figure~\ref{fig:hlum} presents the halo luminosity in molecular hydrogen rotational lines (upper panel), the mass-averaged molecular hydrogen fraction (middle panel), and the fraction of the total $\mathrm{H}_2$ luminosity contributed by the lowest rotational transitions (lower panel) for haloes of different masses at the epochs corresponding to their peak luminosity across redshift $z$.

For low-mass haloes whose virial gas temperature remains below the onset of strong $\mathrm{H}_2$ collisional dissociation ($\sim 3000\text{ K}$), peak emission intensity in the $\mathrm{H}_2$ rotational lines is reached during virialization or at later evolutionary stages as gas density and excitation build up. By contrast, in more massive haloes, maximum luminosity is attained even before full virial equilibrium is established, at epochs when the gas temperature passes through $\sim 3500 - 4000\text{ K}$---where collisional excitation is strong before thermal dissociation and ionization losses become dominant. As shown in the lower panel, the emission is governed by the lowest rotational transitions, and at the epoch of peak luminosity, ortho-$\mathrm{H}_2$ emission dominates over that from para-$\mathrm{H}_2$.

At the epoch of peak rotational line luminosity for a halo of fixed mass $M_{\rm h}$, the mass-averaged molecular hydrogen fraction $f_{\rm H_2}$ exhibits a distinct turnover around a characteristic gas temperature of $T_{\rm gas} \simeq 2000\text{ K}$ as redshift $z$ decreases. In cooler haloes with $T_{\rm gas} < 2000\text{ K}$---typical of low-mass minihalos reaching peak brightness during or after virialization---collisional dissociation is negligible, so longer evolutionary timescales at lower $z$ allow gas-phase formation pathways, predominantly the $\mathrm{H}^-$ channel ($\mathrm{e}^- + \mathrm{H} \rightarrow \mathrm{H}^- + \gamma$ followed by $\mathrm{H}^- + \mathrm{H} \rightarrow \mathrm{H}_2 + \mathrm{e}^-$) and the $\mathrm{H}_2^+$ channel, to accumulate a higher molecular fraction prior to maximum luminosity. Conversely, in warmer haloes with $T_{\rm gas} > 2000\text{ K}$---including heavier minihalos and more massive systems whose peak luminosity occurs as the gas heats toward $\sim 3500 - 4000\text{ K}$ even prior to virial equilibrium---$f_{\rm H_2}$ at peak brightness decreases with decreasing $z$, because stronger excitation of upper rotational and vibrational levels alongside increasingly rapid collisional dissociation ($\mathrm{H}_2 + \mathrm{H} \rightarrow 3\mathrm{H}$ and $\mathrm{H}_2 + \mathrm{e}^- \rightarrow 2\mathrm{H} + \mathrm{e}^-$, intensifying near $T_{\rm diss} \simeq 3350\text{ K}$ and above the ionization threshold $T_{\rm ion} \simeq 1.05 \times 10^4\text{ K}$) depletes $\mathrm{H}_2$, while lower gas densities at smaller $z$ weaken collisional formation relative to destruction.

\section{Modeling $\mathrm{H}_2$ Intensity Maps from Minihalos}
\label{sec:h2_mapping}

To construct three-dimensional intensity maps of molecular hydrogen ($\mathrm{H}_2$) emission during the Dark Ages and Cosmic Dawn, we post-process the spatial overdensity fields and dark matter halo distributions extracted from \texttt{21cmFAST} simulations \citep{Mesinger2007,Mesinger2011} (specifically, coeval boxes and simulated lightcones). Following the theoretical framework for primordial $\mathrm{H}_2$ line formation, our modeling captures thermal rotational emission driven by collisional excitation within minihalos across eight rotational transitions: $2\to0$, $3\to1$, $4\to2$, $5\to3$, $6\to4$, $7\to5$, $8\to6$, and $9\to7$.

\subsection{Mechanisms of Line Formation and Single-Halo Luminosity}
\label{subsec:line_mechanisms}

In the low-density regime characteristic of primordial minihalos, rotational level populations are governed by a non-LTE balance between collisional excitations (primarily with neutral hydrogen and helium) and spontaneous radiative decay. The total thermal luminosity $L_{\mathrm{th}}$ for a specific rotational transition ($J \to J-2$) integrated over the halo volume is expressed as:
\begin{equation}
    L_{\mathrm{th}}^{(i)}(M, z) = \int \Lambda_{\mathrm{H}_2}^{(i)}(T) \, n_{\mathrm{H}} \, n_{\mathrm{H}_2} \, dV \approx \eta_i(M, z) \, L_{\mathrm{H}_2}(M, z),
\end{equation}
where $L_{\mathrm{H}_2}(M, z)$ is the total bolometric $\mathrm{H}_2$ cooling luminosity of a halo of mass $M$ at redshift $z$, $\Lambda_{\mathrm{H}_2}^{(i)}(T)$ is the temperature-dependent cooling function for the $i$-th rotational line, and $\eta_i(M, z)$ represents the fractional luminosity emitted in that transition (incorporating the equilibrium ortho-to-para statistical weight ratio of $3:1$). In our numerical pipeline, $L_{\mathrm{H}_2}(M, z)$ and the transition fractions $\eta_i(M, z)$ are evaluated in SI units ($\mathrm{W}$) using a two-dimensional thin-plate spline radial basis function (RBF) interpolator calibrated against detailed halo cooling models across the $(M, z)$ parameter space.

\subsection{Spatial Distribution and Number Density Maps of Minihalos}
\label{subsec:minihalo_maps}

\begin{figure}
   \centering
   \includegraphics[width=0.48\textwidth]{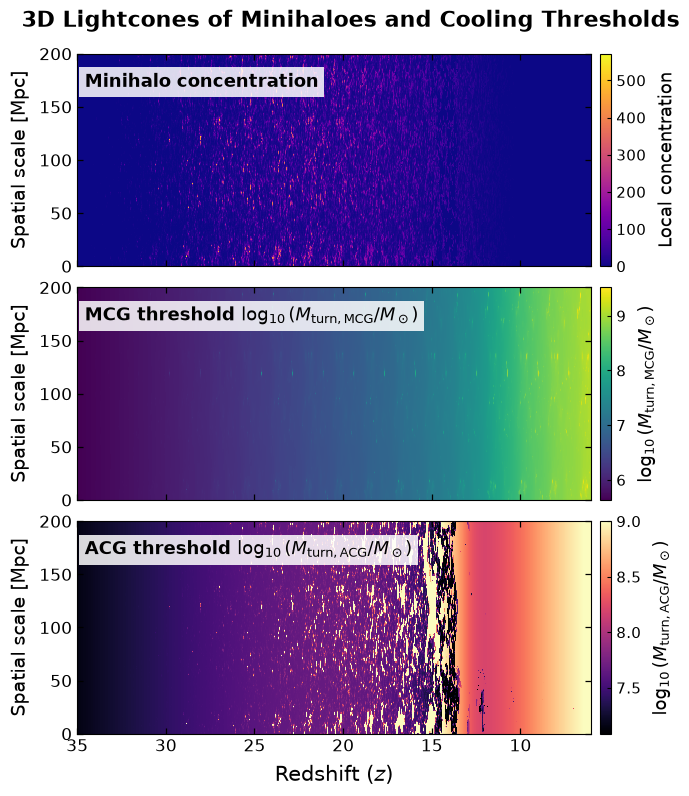}
   \caption{Spatial slices along the simulated lightcone illustrating the 3D evolution of local minihalo number density $n_{\mathrm{halo}}$ (top), molecular cooling galaxy (MCG) turnover mass threshold $M_{\mathrm{turn, MCG}}$ (middle), and atomic cooling galaxy (ACG) turnover mass threshold $M_{\mathrm{turn, ACG}}$ (bottom) as a function of redshift $z$ and transverse spatial scale ($\mathrm{Mpc}$). The horizontal axis is oriented to follow cosmic evolution from high to low redshift.}
   \label{fig:lightcone_minihalo_evolution}
\end{figure}
\begin{figure}
   \centering
   \includegraphics[width=0.48\textwidth]{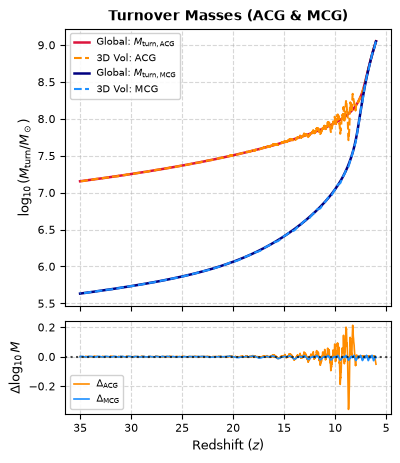}
   \caption{Redshift evolution of the turnover masses ($\log_{10} M_{\mathrm{turn}}$) for atomic cooling galaxies (ACG, red/orange) and molecular cooling galaxies (MCG, navy/blue). The top panel compares global analytical tracks (solid lines) with 3D volume averages extracted from the lightcone simulation (dashed lines). The bottom panel displays the corresponding residual deviations $\Delta \log_{10} M$ relative to the global models.}
   \label{fig:mturn_acg_mcg_comparison}
\end{figure}

To map the spatial abundance of active minihalos across the cosmic web, we compute their local comoving number density $n_{\mathrm{halo}}(\mathbf{x}, z)$ (in $\mathrm{cMpc}^{-3}$) within each simulation grid voxel at comoving position $\mathbf{x}$. Because individual minihalos ($M \sim 10^5\text{--}10^7\,M_\odot$) fall below the direct $N$-body particle resolution of large-scale cosmological volumes, their sub-grid distribution is modeled analytically using the conditional halo mass function (CHMF) derived from the Sheth--Tormen ellipsoidal collapse formalism, conditioned on the local matter overdensity $\delta_{\mathrm{vox}}(\mathbf{x}, z)$:
\begin{equation}
    n_{\mathrm{halo}}(\mathbf{x}, z) = \int_{\ln M_{\mathrm{min}}(\mathbf{x}, z)}^{\ln M_{\mathrm{max}}(\mathbf{x}, z)} \frac{dn}{d\ln M}\left(M \Big| \delta_{\mathrm{vox}}(\mathbf{x}, z), z\right) d\ln M,
\end{equation}
with the total number of minihalos per voxel given by $N_{\mathrm{halo}}(\mathbf{x}, z) = n_{\mathrm{halo}}(\mathbf{x}, z) \, V_{\mathrm{voxel}}$, where $V_{\mathrm{voxel}} = (L_{\mathrm{box}} / N_{\mathrm{grid}})^3$ is the comoving voxel volume.

As illustrated in Figure~\ref{fig:lightcone_minihalo_evolution}, the effective mass window $[M_{\mathrm{min}}, M_{\mathrm{max}}]$ is regulated inhomogeneously across space and time by two competing radiative feedback channels:
\begin{itemize}
    \item \textbf{Lower Mass Limit ($M_{\mathrm{min}}$):} Set by the local molecular cooling turnover mass $M_{\mathrm{turn, MCG}}(\mathbf{x}, z) = M_{\mathrm{filter}}(z)\left[1 + A_{\mathrm{LW}} J_{21}(\mathbf{x}, z)^{\beta_{\mathrm{LW}}}\right]$, which accounts for baryon--dark matter streaming velocities and $\mathrm{H}_2$ photo-dissociation driven by the local Lyman--Werner (LW) radiation intensity $J_{21}(\mathbf{x}, z)$.
    \item \textbf{Upper Mass Limit ($M_{\mathrm{max}}$):} Bounded by the atomic cooling turnover mass $M_{\mathrm{turn, ACG}}(\mathbf{x}, z)$ ($T_{\mathrm{vir}} \approx 10^4\ \mathrm{K}$), which shifts upward in response to local photoionization heating parameterized by the ionized fraction field $x_{\mathrm{HII}}(\mathbf{x}, z) = 1 - x_{\mathrm{HI}}(\mathbf{x}, z)$. In strongly irradiated or ionized regions where feedback drives $M_{\mathrm{min}}(\mathbf{x}, z) \ge M_{\mathrm{max}}(\mathbf{x}, z)$, minihalo star formation and cooling are fully suppressed ($n_{\mathrm{halo}} = 0$).
\end{itemize}

Numerically, the CHMF integral is evaluated across the lightcone via vectorized interpolation over precomputed cumulative mass profiles generated by internal C-routines (\texttt{cfuncs.integrate\_chmf\_interval}) on a discretized grid of overdensities $\delta$ and logarithmic mass nodes $\ln M \in [\ln(10^4\,M_\odot), \ln(10^9\,M_\odot)]$.

Figure~\ref{fig:mturn_acg_mcg_comparison} compares the global analytical turnover mass tracks against the 3D volume-averaged thresholds extracted from our lightcone simulation, highlighting the spatial variance induced by inhomogeneous LW and ionization feedback.

\subsection{Voxel Luminosity Density and Specific Intensity Calculation}
\label{subsec:voxel_intensity}

\begin{figure}
   \centering
   \includegraphics[width=0.48\textwidth]{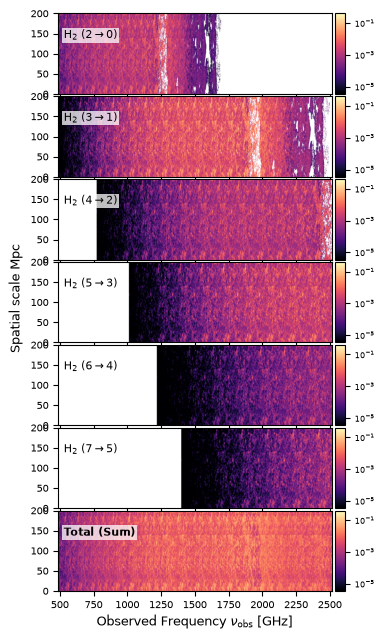}
   \caption{Single-column multi-panel visualization of 3D lightcone intensity mapping slices for six individual $\mathrm{H}_2$ rotational transitions  and their total co-added intensity map (bottom panel). The horizontal axis corresponds to the observed frequency $\nu_{\mathrm{obs}}$ [GHz], and the vertical axis indicates the transverse spatial scale in Mpc.}
   \label{fig:multiline_lightcone_slices}
\end{figure}

\begin{figure}
   \centering
   \includegraphics[width=0.48\textwidth]{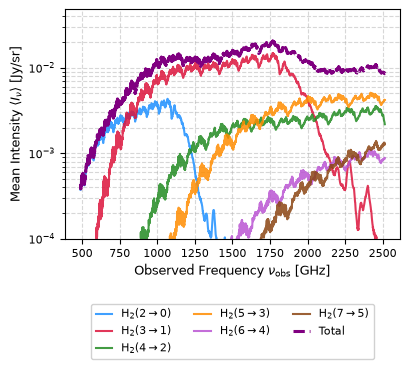}
   \caption{Evolution of the global mean specific intensity $\langle I_\nu \rangle$ as a function of observed frequency $\nu_{\mathrm{obs}}$. Colored solid curves represent individual contributions from the six $\mathrm{H}_2$ rotational transitions ($2\to0$ through $7\to5$), while the dashed purple curve shows the total co-added background signal.}
   \label{fig:global_minihalo_intensity_evolution}
\end{figure}

Combining the spatially varying minihalo mass distribution with the single-halo cooling model, we compute the comoving luminosity density $\rho_L^{(i)}(\mathbf{x}, z)$ for each rotational transition $i$ by weighting the CHMF with the transition-specific luminosity $L_{\mathrm{th}}^{(i)}(M, z)$:
\begin{equation}
    \rho_L^{(i)}(\mathbf{x}, z) = \int_{\ln M_{\mathrm{min}}(\mathbf{x}, z)}^{\ln M_{\mathrm{max}}(\mathbf{x}, z)} L_{\mathrm{th}}^{(i)}(M, z) \, \frac{dn}{d\ln M}\left(M \Big| \delta_{\mathrm{vox}}(\mathbf{x}, z), z\right) d\ln M,
\end{equation}
where $\rho_L^{(i)}(\mathbf{x}, z) = 0$ in sterilized voxels ($M_{\mathrm{min}} \ge M_{\mathrm{max}}$). In the discretized pipeline, this integral is evaluated simultaneously with $n_{\mathrm{halo}}$ by accumulating differential luminosity contributions $\Delta \rho_L^{(i)} = \Delta n_{\mathrm{halo}} \, \langle L_{\mathrm{th}}^{(i)} \rangle_{\mathrm{bin}}$ across the mass bins and interpolating the resulting cumulative luminosity profiles at $\ln M_{\mathrm{min}}$ and $\ln M_{\mathrm{max}}$.

For a given rotational transition with rest-frame frequency $\nu_{0, i} = c / \lambda_i$, the observable specific intensity $I_{\nu}^{(i)}(\mathbf{x}, z)$ (in $\mathrm{Jy\,sr^{-1}}$) is computed from the comoving luminosity density via the standard cosmological radiative transfer relation:
\begin{equation}
    I_{\nu}^{(i)}(\mathbf{x}, z) = \frac{c}{4\pi H(z) \nu_{0, i}} \, \rho_L^{(i)}(\mathbf{x}, z) \times 10^{26},
\end{equation}
where $c$ is the speed of light, $H(z)$ is the Hubble parameter at redshift $z$, $\rho_L^{(i)}$ is converted to $\mathrm{W\,m^{-3}}$, and the factor $10^{26}$ converts spectral radiance into Janskys per steradian ($1\,\mathrm{Jy} = 10^{-26}\,\mathrm{W\,m^{-2}\,Hz^{-1}}$).

To model multi-line intensity mapping across observed frequencies, individual transition cubes are projected onto a unified frequency grid referenced to the $\mathrm{H}_2\ 3\to1$ rest frame ($\nu_{\mathrm{obs}} = \nu_{0, 31} / (1+z)$). Figures~\ref{fig:multiline_lightcone_slices} and \ref{fig:global_minihalo_intensity_evolution} present the resulting multi-line 3D lightcone slices and the redshift/frequency evolution of the global mean specific intensity $\langle I_\nu \rangle$, illustrating both individual transition contributions and their co-added signal. The frequency dependence of the global $\mathrm{H}_2$ line-intensity
spectrum reveals distinct contributions from individual rotational
transitions. The ortho-$\mathrm{H}_2$ $J=3\rightarrow1$ line
($\lambda_0=17.03\ \mu\mathrm{m}$) provides the largest contribution
and dominates the spectrum over the observed frequency range
$0.8$--$1.9\ \mathrm{THz}$. Below $0.8\ \mathrm{THz}$, the
para-$\mathrm{H}_2$ $J=2\rightarrow0$ transition
($\lambda_0=28.2\ \mu\mathrm{m}$) becomes dominant, whereas at
frequencies above $1.9\ \mathrm{THz}$ the spectrum is mainly
determined by the ortho-$\mathrm{H}_2$ $J=5\rightarrow3$ and
para-$\mathrm{H}_2$ $J=4\rightarrow2$ transitions. This distinct
frequency-dependent dominance of individual lines demonstrates
the complementary contributions of different rotational transitions
to the global molecular emission.

\begin{figure}
   \centering
   \includegraphics[width=0.48\textwidth]{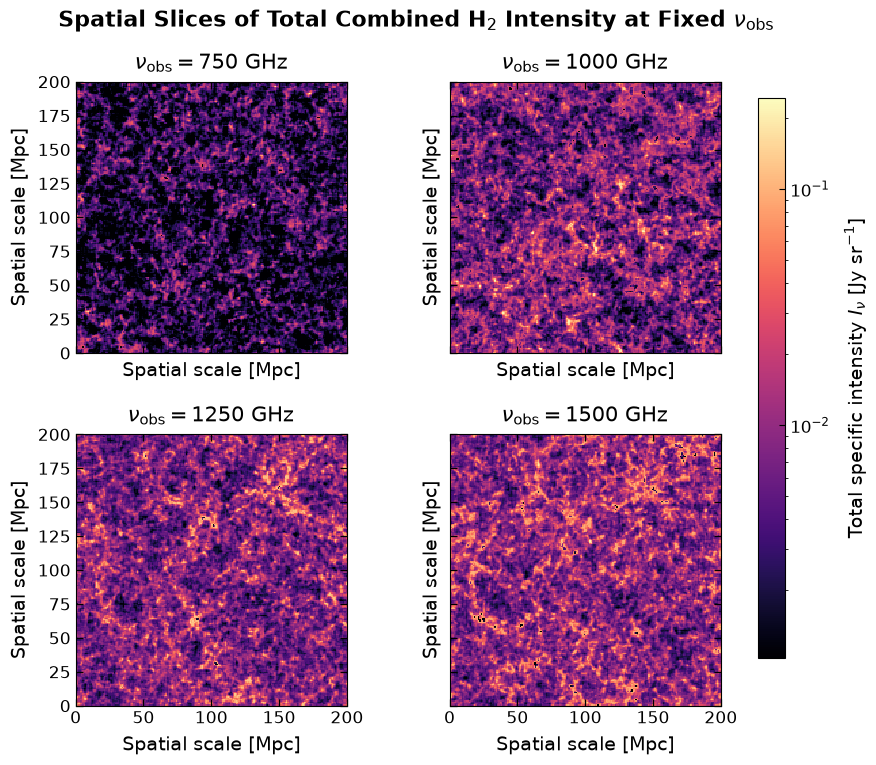}
   \caption{Two-dimensional spatial slices along the simulated lightcone showing the total co-added specific intensity $I_{\nu}$ from lowest six $\mathrm{H}_2$ rotational transitions, evaluated at four fixed observed frequencies: $\nu_{\mathrm{obs}} = 1000$~GHz (top left), $1500$~GHz (top right), $2000$~GHz (bottom left), and $2500$~GHz (bottom right). The intensity values are displayed on a logarithmic scale ($\mathrm{Jy\,sr^{-1}}$), with transverse spatial scales expressed in comoving Mpc.}
   \label{fig:intensity_spatial_slices}
\end{figure}

Finally, Figure~\ref{fig:intensity_spatial_slices} illustrates the two-dimensional spatial morphology of the co-added $\mathrm{H}_2$ intensity maps across fixed observed frequency slices ($\nu_{\mathrm{obs}} = 1000, 1500, 2000$, and $2500\ \mathrm{GHz}$), capturing the strongly clustered filamentary distribution of minihalo emission across cosmological volumes.

\section{Discussion and conclusions}
\label{sec:conclusions}

In this work, we have performed comprehensive numerical modelling of
primordial halo luminosities and three-dimensional intensity
fluctuations in the rotational emission lines of molecular hydrogen
($\mathrm{H}_2$) throughout the Dark Ages and Cosmic Dawn
($6 \leq z \leq 35$).

We have developed a self-consistent framework for describing the
chemical, thermal, and rotational excitation evolution of primordial
molecular hydrogen in the cosmological background and collapsing
density perturbations. The chemical model is based on an extended
reaction network comprising 20 atomic, ionic, and molecular species
and 166 chemical reactions. We have also formulated a kinetic
description of the rotational level populations of $\mathrm{H}_2$,
accounting for radiative transitions induced by the cosmic microwave
background (CMB), collisions with the main constituents of primordial
gas, and the formation and destruction of molecules. This approach
allows us to follow the non-equilibrium evolution of the rotational
populations and the relative abundances of ortho- and para-$\mathrm{H}_2$.

The thermal evolution model incorporates molecular and chemical
heating and cooling processes, as well as the transition of collapsing
perturbations to the virialized state. Our results demonstrate that
molecular formation, rotational excitation, and the thermal evolution
of the gas are closely coupled. In particular, collisional dissociation
can substantially reduce the molecular abundance in sufficiently hot
halos, thereby affecting their line emission.

We have calculated the intrinsic luminosities of individual
$\mathrm{H}_2$ rotational transitions and their fractional
contributions to the total rotational emission. The results show that
the lowest rotational transitions dominate the total line luminosity,
with ortho-$\mathrm{H}_2$ generally providing the largest contribution
at the epoch of peak halo emission. The relative contributions of
individual transitions depend on halo mass, collapse redshift, and
the evolutionary stage of the gas. We have identified two distinct
regimes of maximum emission: in low-mass halos, the luminosity peaks
during or after virialization, whereas in more massive halos it
reaches its maximum during the heating phase, before the establishment
of full virial equilibrium. These results highlight the importance of
a consistent treatment of chemical evolution, thermal balance, and
rotational excitation when predicting the molecular emission of
primordial halos.

A comparison with the results of \citet{Gong2013} reveals that their
predicted halo intensities in $\mathrm{H}_2$ rotational lines are
approximately three orders of magnitude higher than those obtained
in our calculations. Moreover, the peak of their spectral intensity
distribution occurs at higher rotational quantum numbers $J$.
A possible explanation for this discrepancy is the treatment of the
inhomogeneous radial structure of halos in their model, which
incorporates spatial variations in gas density, temperature, and
molecular abundance. In particular, the $\mathrm{H}_2$ abundance in
the central regions of their halos is several orders of magnitude
higher than in the outer regions. At sufficiently high central
densities, three-body reactions may also become efficient, further
enhancing molecular hydrogen formation. Consequently, the central
regions can make a dominant contribution to the integrated halo
luminosity, potentially increasing it substantially compared with
models based on homogeneous gas distributions.

Another possible source of the difference in the predicted line
spectra is the treatment of rotational excitation. The Boltzmann
prescription adopted by \citet{Gong2013} may overestimate the
populations of high-$J$ levels in regions where collisional processes
are insufficient to establish local thermodynamic equilibrium (LTE).
This could contribute to shifting the spectral intensity peak towards
higher rotational transitions. Thus, although an inhomogeneous
description of the halo gas is physically more realistic, reliable
predictions of individual line luminosities require a consistent
treatment of non-LTE rotational populations and, where necessary,
radiative transfer effects.

To investigate the contribution of the minihalo population to the
large-scale molecular emission, we have used the cosmological
simulation code \texttt{21cmFAST} to generate three-dimensional
distributions of minihalo number density and $\mathrm{H}_2$ line
intensity fluctuations in simulation volumes of
$(200\ \mathrm{Mpc})^3$, with a spatial resolution of $1\ \mathrm{Mpc}$.
By integrating the conditional halo mass function, we have accounted
for the spatially inhomogeneous effects of radiative feedback.
In particular, the local Lyman--Werner background
($J_{\mathrm{LW}}$) determines the lower mass threshold for minihalo
formation, whereas the local ionized fraction ($x_{\mathrm{HII}}$)
sets the upper mass limit of the minihalo population.

The resulting global $\mathrm{H}_2$ line-intensity spectrum exhibits
a broad maximum near an observed frequency of
$\nu_{\mathrm{obs}}\sim 2\ \mathrm{THz}$, resembling the spectral
shape of the Cosmic Infrared Background (CIB) \citep{Fixsen_1998,Odegard_2019}. However, within the adopted homogeneous-halo
approximation, the predicted $\mathrm{H}_2$ signal is approximately
eight orders of magnitude fainter than the CIB. This low amplitude
partly reflects the limited molecular abundance resulting from the
inefficiency of three-body $\mathrm{H}_2$ formation at the densities
considered in the model. A more realistic treatment of the internal
density structure of collapsing minihalos could substantially
increase their molecular abundances and, consequently, their
integrated line emission.

The contributions of individual rotational transitions to the
global spectrum vary significantly with observed frequency.
Among the transitions considered, the ortho-$\mathrm{H}_2$
$J=3\rightarrow1$ line ($\lambda_0=17.03\ \mu\mathrm{m}$)
has the highest luminosity and dominates the emission over the
frequency range of approximately $0.8$--$1.9\ \mathrm{THz}$.
At frequencies below $0.8\ \mathrm{THz}$, the para-$\mathrm{H}_2$
$J=2\rightarrow0$ line ($\lambda_0=28.2\ \mu\mathrm{m}$)
provides the dominant contribution. At frequencies above
$1.9\ \mathrm{THz}$, the emission is mainly contributed by the
ortho-$\mathrm{H}_2$ $J=5\rightarrow3$ and para-$\mathrm{H}_2$
$J=4\rightarrow2$ transitions. Thus, different rotational lines
dominate in different observed frequency intervals, providing
complementary spectral windows for investigating the primordial
molecular emission.

Our estimates suggest that an enhancement of the $\mathrm{H}_2$
signal by at least two orders of magnitude could bring it closer to
the sensitivity range of next-generation space-based far-infrared
and submillimetre spectrometers, including the proposed FOSSIL
mission \citep{Aghanim2026FOSSIL}. This possibility, however, needs to be assessed using
models that consistently account for the internal gas structure,
non-equilibrium molecular excitation, and radiative transfer.

The frequency-dependent contributions of these rotational
transitions provide an opportunity to develop a multi-line
spectral tomography approach to studying the early Universe.
In particular, the simultaneous observation of the ortho-$\mathrm{H}_2$
$J=3\rightarrow1$ and $J=5\rightarrow3$ lines and the
para-$\mathrm{H}_2$ $J=2\rightarrow0$ and $J=4\rightarrow2$
lines would allow intensity maps obtained at different observed
frequencies to be cross-correlated. Since emission from the same
cosmological redshift is observed at different frequencies in
different transitions, such cross-correlations could reveal common
spatial structures associated with the same underlying population
of primordial minihalos. This approach could also help distinguish
the primordial molecular signal from foreground emission and
improve the reliability of its identification. By extending molecular
intensity mapping beyond measurements of the global spectrum,
multi-line spectral tomography could potentially enable the
three-dimensional reconstruction of the distribution of the first
gravitationally bound structures. The results presented here provide
a theoretical basis for developing such techniques and for assessing
the spectral coverage, sensitivity, and angular and frequency
resolution required for future observations of primordial molecular
emission.

\begin{acknowledgements}
 This work is done in the framework of the project
  \emph{''Tomography of the Dark Ages and Cosmic Dawn in the lines of
    hydrogen and the first molecules as a test of cosmological
    models''} (state registration number 0124U004029) supported by
  National Research Found of Ukraine.
  This research was funded partly by the Research Council of Norway, project number 362008. Authors  thank Prof. Andrei Mesinger and his team for fruitful discussions and technical assistance with the 21cmFAST simulations. YK would like to thank the Scuola Normale Superiore in Pisa for its warm hospitality. The authors used Gemini (Google) and ChatGPT (OpenAI) to improve the language and clarity of the manuscript and to assist in developing post-processing scripts. All AI-assisted text and code were carefully reviewed and verified by the authors, who take full responsibility for the content of this publication.
\end{acknowledgements}

\bibliographystyle{aa}
\bibliography{refs}

\onecolumn

\begin{appendix}

\end{appendix}

\end{document}